\documentclass[twocolumn,pra,aps,showpacs,floatfix]{revtex4}
\usepackage[latin9]{inputenc}
\usepackage{amsmath}
\usepackage{amssymb}
\usepackage{graphicx}
\usepackage{esint}
\usepackage{amsfonts}
\usepackage{eurosym}
\usepackage{color}
\usepackage{xcolor}
\usepackage{mathrsfs}
\usepackage{amsmath}
\usepackage{graphicx}
\usepackage{dcolumn}
\usepackage{bm}
\usepackage{times}
\usepackage{amssymb}
\usepackage{float}
\usepackage{array}
\usepackage{tikz}

\makeatletter
\@ifundefined{textcolor}{}
{
 \definecolor{BLACK}{gray}{0}
 \definecolor{WHITE}{gray}{1}
 \definecolor{RED}{rgb}{1,0,0}
 \definecolor{GREEN}{rgb}{0,1,0}
 \definecolor{BLUE}{rgb}{0,0,1}
 \definecolor{CYAN}{cmyk}{1,0,0,0}
 \definecolor{MAGENTA}{cmyk}{0,1,0,0}
 \definecolor{YELLOW}{cmyk}{0,0,1,0}
}
\@ifundefined{textcolor}{}{ \definecolor{BLACK}{gray}{0}
 \definecolor{WHITE}{gray}{1}
 \definecolor{RED}{rgb}{1,0,0}
 \definecolor{GREEN}{rgb}{0,1,0}
 \definecolor{BLUE}{rgb}{0,0,1}
 \definecolor{CYAN}{cmyk}{1,0,0,0}
 \definecolor{MAGENTA}{cmyk}{0,1,0,0}
 \definecolor{YELLOW}{cmyk}{0,0,1,0}
}

\makeatother

\begin{document}

\title{Phase tunable Josephson junction and spontaneous mass current in a
spin-orbit coupled Fermi superfluid}
\author{Lei Jiang$^{1}$}
\author{Yong Xu$^{1,2}$}
\author{Chuanwei Zhang$^{1}$}
\thanks{Corresponding Author, Email: chuanwei.zhang@utdallas.edu}
\affiliation{$^{1}$Department of Physics, The University of Texas at Dallas, Richardson,
Texas 75080, USA\\
$^{2}$Department of Physics, University of Michigan, Ann Arbor, Michigan
48109, USA}

\begin{abstract}
Atomtronics has the potential for engineering new types of functional
devices, such as Josephson junctions (JJs). Previous studies have mainly
focused on JJs whose ground states have 0 or $\pi $ superconducting phase
difference across the junctions, while arbitrarily tunable phase JJs may
have important applications in superconducting electronics and quantum
computation. Here we show that a phase tunable JJ can be implemented in a
spin-orbit coupled cold atomic gas with the magnetic tunneling barrier
generated by a spin-dependent focused laser beam. We consider the JJ
confined in either a linear harmonic trap or a circular ring trap. In the
ring trap, the magnetic barrier induces a spontaneous mass current for the
ground state of the JJ, demonstrating the magnetoelectric effects of cold
atoms.
\end{abstract}

\maketitle

Atomtronics is a new exciting interdisciplinary field \cite%
{zoller04,holland07,holland09,ott15} aiming to mimic electronic circuits and
build new functional devices, utilizing the high controllability and purity
of cold atomic gases. Recently, atomic Josephson junctions (JJs) have been
realized \cite{Ryu13PRL,Campbell13,Campbell14} in toroidal Bose-Einstein
Condensates (BECs) \cite{Stamper-Kurn05,Phillips07,Boshier09,Campbell11},
analogous to the well-known superconducting quantum interference devices
(SQUID). In solid-state devices, besides the common zero phase, the ground
state of a JJ may possess a $\pi $ phase of the superconducting order
parameter across the junction \cite{Bulaevskii77}, which can be generated by
inserting a layer of insulator with magnetic impurities \cite{Vavra06} or a
layer of ferromagnetic material \cite{Ryazanov01} or an unconventional
superconductor \cite{Barone87,Tsuei00} between two regular \textit{s}-wave
superconductors. Such $\pi $ phase JJs resemble the Larkin--{}Ovchinnikov
(LO) state in a spin imbalanced superconductor \cite{lo65} and have also
been studied in cold atomic gases \cite{kulic07,ohashi10}. More generally,
the phase of the JJ ground state could be arbitrary (not 0 or $\pi $), which
may have important applications such as phase batteries and rectifiers \cite%
{Reynoso08,Reynoso12}, phase-based quantum bits \cite{Padurariu10}, \textit{%
etc}. Recently, such arbitrary phase JJs have been experimentally realized
using nanowire quantum dots \cite{Kouwenhoven}. However, arbitrary phase JJs
have not been explored in atomtronics.

In cold atomic gases, synthetic gauge fields and spin-orbit coupling have
paved a way for neutral atoms to interact with external synthetic electric
and magnetic fields \cite%
{Spielman09,Spielman11a,Spielman11b,pan12,engels13,chen14,zhang12,cheuk12,Spielman13b}%
. In particular, the 1D equal-Rashba-Dresselhaus spin-orbital coupling has
been realized experimentally for fermions using a two-photon Raman process
\cite{zhang12,cheuk12,Spielman13b}. Interestingly, in the presence of the
Raman detuning which acts as an in-plane Zeeman field, the inversion
symmetry of the Fermi surface is broken, leading to the Fulde-Ferrell (FF)
superfluid \cite{ff64} with a spatially modulating phase of the order
parameter \cite{zheng13,yi12,qu13,yi13,hu13,chen13,pu13,xu14}. Therefore a
nature question is whether such spatially modulating phase of the FF state
could be used to engineer JJs with arbitrary and tunable phases.

\begin{figure}[b]
\centering
\includegraphics[width=3.2in]{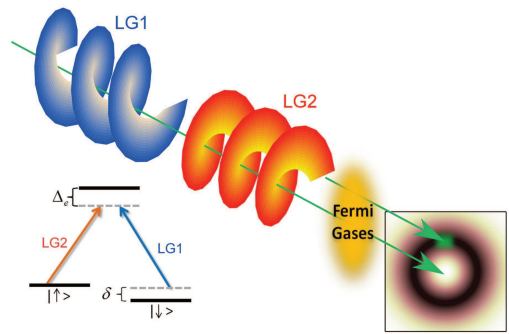}
\caption{(Color online) Illustration of the proposed experimental setup. Two
copropagating LG beams (LG1 and LG2) with different orbital angular momenta
couple two atomic hyperfine states in the ring structure to induce SOAM
coupling through the Raman process. An additional spin-dependent focused
laser beam provides a local magnetic barrier (the green spot).}
\label{f0}
\end{figure}

In this article, we address this issue by studying a JJ generated by a
magnetic barrier in a 1D spin-orbit coupled Fermi superfluid. We consider
two types of traps: a linear harmonic trap and a circular ring trap. For the
former, the spin-momentum coupling has been experimentally realized for
Fermi gases \cite{zhang12,cheuk12,Spielman13b}, and for the latter, the
corresponding spin-orbital-angular-momentum (SOAM) coupling was proposed
\cite{Hu15,pu15,sun15,qu15} to be realized using Laguerre-Gaussian (LG)
laser beams \cite{Wright97,Phillips06,Bigelow09,Hadzibabic13} (see Fig. \ref%
{f0}). The magnetic barrier for the JJ can be generated by a spin-dependent
focused laser beam. In both types of traps, the phase across the JJ can be
continuously tuned by changing the parameters of the magnetic barrier.
Interestingly, we find that the magnetic barrier induces a spontaneous
finite mass current for the ground state of the JJ in a ring trap, which
realizes the magnetoelectric effects in atomtronics.

\textit{Model}: For simplicity of the numerical calculation, hereafter we
consider 1D Fermi gases, but the results apply to 2D and 3D due to the same
mechanism for generating the FF phase junction. We first consider a
spin-momentum coupled Fermi gas confined in a 1D harmonic trap. Within the
mean-field approximation, the dynamics of the system is governed by the
mean-field many-body Hamiltonian
\begin{equation}
H=\int dx\left\{ \hat{\Psi}^{+}H_{S}\hat{\Psi}-\left[ \Delta (x)\hat{\psi}%
_{\uparrow }^{+}\hat{\psi}_{\downarrow }^{+}+h.c.\right] \right\} ,
\end{equation}%
where $\hat{\Psi}=(\hat{\psi}_{\uparrow }(x),\,\hat{\psi}_{\downarrow
}(x))^{T}$ and $\hat{\psi}_{\sigma }(x)$ with $\sigma =\uparrow ,\downarrow $
are fermionic annihilation operators for the spin $\sigma $ state. The
single-particle Hamiltonian $H_{S}=H_{0}+H_{\mathrm{SOC}}+H_{Z}$. $H_{0}=[-%
\frac{\hbar ^{2}\partial ^{2}}{2m\partial x^{2}}-\mu +\frac{m\omega ^{2}x^{2}%
}{2}]$ with the Planck constant $\hbar $, the mass of atoms $m$, the
chemical potential $\mu $, and the harmonic trap frequency $\omega $. The
spin-momentum coupling term $H_{\mathrm{SOC}}=-i\lambda \sigma _{z}\partial
_{x}$ with Pauli matrices $\sigma _{i=x,y,z}$ and coupling strength $\lambda
=\hbar ^{2}k_{R}/m$, where $k_{R}$ is the recoil momentum of the Raman
laser. The Zeeman field term $H_{Z}=-\Omega _{R}\sigma _{x}+V_{Z}(x)\sigma
_{z}$ with $\Omega _{R}$ and $V_{Z}(x)$ being the out-of-plane and in-plane
Zeeman field strengths. $\Omega _{R}$ is determined by the Raman laser
intensities and $V_{Z}(x)$ is induced by the local magnetic barrier. The
order parameter $\Delta (x)\equiv -g_{1D}\langle \hat{\psi}_{\downarrow }(x)%
\hat{\psi}_{\uparrow }(x)\rangle $. The constant $g_{1D}$ is the 1D two-body
$s$-wave interaction strength, which can be characterized by a scaleless
parameter $\gamma \equiv -mg_{1D}/(\hbar ^{2}n_{0})$ that represents the
ratio between the interaction and kinetic energy. Here $n_{0}=(2/\pi )\sqrt{%
Nm\omega /\hbar }$ with $N$ being the total number of atoms.

In terms of the Nambu spinor $\hat{\Phi}(x)=[\hat{\psi}_{\uparrow }(x),\hat{%
\psi}_{\downarrow }(x),\hat{\psi}_{\uparrow }^{+}(x),\hat{\psi}_{\downarrow
}^{+}(x)]^{T}$, the mean-field Hamiltonian $H=\frac{1}{2}\int dx\,\hat{\Phi}%
^{+}(x)H_{\mathrm{BdG}}\hat{\Phi}(x)$ can be numerically solved using the
hybrid self-consistent Bogoliubov-de Gennes (BdG) method~\cite%
{liu07,liu13d,jiang11}. The BdG quasi-particles are obtained by
diagonalizing
\begin{equation}
H_{\mathrm{BdG}}\,\varphi _{\eta }(x)=E_{\eta }\,\varphi _{\eta }(x)\,,
\label{bdg}
\end{equation}%
with energies $E_{\eta }$ and wavefunctions $\varphi _{\eta
}(x)=[u_{\uparrow \eta }(x),u_{\downarrow \eta }(x),v_{\uparrow \eta
}(x),v_{\downarrow \eta }(x)]^{T}$ indexed by subscript $\eta =1,2,3\ldots $
The wavefunctions are normalized such that ${\sum_{\sigma =\uparrow
,\downarrow }}\int dx(|u_{\sigma \eta }(x)|^{2}+|v_{\sigma \eta }(x)|^{2})=1$%
.

We use a \textquotedblleft hybrid\textquotedblright\ method of Ref. \cite%
{liu07,liu13d} to solve the eigenvalue problem of Eq. (\ref{bdg}). We get
all eigenenergy pairs of $H_{\mathrm{BdG}}$ with energy $|E|\leqslant E_{c}$%
, where $E_{c}$ is a cut-off energy that is chosen to be large compared to
the Fermi energy but small compared to the width of the discretized $H_{%
\mathrm{BdG}}$ spectral. Typically we take $E_{c}=8E_{F}$ with the
non-interacting Fermi energy $E_{F}=\hbar \omega {N}/{2}$ in a harmonic
trap. The Fermi wave number $k_{F}$ is obtained from $E_{F}=\hbar
^{2}k_{F}^{2}/2m$, and the Thomas-Fermi radius $x_{\mathrm{TF}}=\sqrt{N\hbar
/(m\omega )}$. For this eigenstate problem, we use the discrete variable
representation (DVR) of the plane wave basis \cite{Colber92}. The order
parameter is
\begin{equation}
\Delta (x)=-\frac{g_{1D}}{2}\underset{\eta }{\sum }\,[u_{\uparrow \eta
}v_{\downarrow \eta }^{\ast }f(E_{\eta })+u_{\downarrow \eta }v_{\uparrow
\eta }^{\ast }f(-E_{\eta })]\,,
\end{equation}%
where  $f(E)=1/[e^{E/k_{B}T}+1]$ is Fermi-Dirac distribution function and $T$
is the temperature. Here we present results for $T=0$. For states above the
energy cut-off $E_{c}$, we employ a semi-classical method based on the local
density approximation. The new order parameter is calculated by combining
the contributions from the DVR and semi-classical solutions and is put back
to the mean-field Hamiltonian. The procedure is repeated until the order
parameter converges.

In the ring trap, the corresponding SOAM coupling is realized by two LG
Raman lasers with opposite OAM ($L_{1}=-L_{2}=L$)~\cite{Hu15,pu15,sun15,qu15}
(Fig.~\ref{f0}). The Hamiltonian is similar except there is no harmonic trap
and $x$ is changed to $R\theta $, where $R$ is the radius of the ring. The
azimuth angle $\theta $ is in the range $\theta \in \lbrack -\pi ,\,\pi ]$.
The SOAM coupling strength is $\lambda =\hbar ^{2}L/mR$.

\begin{figure}[tbp]
\includegraphics[width=0.5\textwidth]{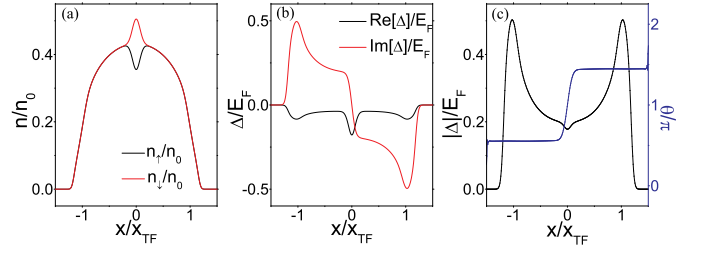}
\caption{Tunable phase Josephson junctions in a 1D harmonic trap. The
magnetic barrier is located at the center of the harmonic trap. (a) Atomic
density profiles. (b) The real and imaginary parts of the order parameter.
(c) The absolute value and phase of the order parameter. $%
u_{mag}=0.05E_{F}x_{TF}$, $a_{imp}=0.1x_{TF}$,$\protect\gamma =2$.2, $%
\protect\lambda =1.5E_{F}/k_{F}$, $\Omega _{R}=0.8E_{F}$, $\protect\mu %
=0.285E_{F}$, $N=60$.}
\label{f1}
\end{figure}

\emph{Phase tunable Josephson junction}: We first consider a magnetic
barrier for the JJ located at center of the harmonic trap and generated by a
spin-dependent focused laser beam with $V_{Z}(x)=u_{mag}e^{-(\frac{x}{a_{mag}%
})^{2}}/a_{mag}\sqrt{\pi }$, where $u_{mag}>0$ and $a_{mag}$ are barrier
strength and width respectively. In the absence of the magnetic barrier, the
system is population balanced and the density exhibits a parabolic profile
which can be described by\ the Thomas-Fermi approximation. In the presence
of a local magnetic barrier, the density exhibits a dip for spin $\uparrow $
atoms and a bump for spin $\downarrow $ ones at the center of the trap [as
shown in Fig.~\ref{f1}(a)] because of opposite potentials for two spins.

The magnetic barrier acts as a local in-plane Zeeman field and induces the
local FF type of order parameter, as shown in Fig.~\ref{f1}(b) where both
real and imaginary parts of the order parameter are non-zero. In Fig.~\ref%
{f1}(c), we plot the absolute value of the order parameter, showing a small
dip inside the barrier due to the suppression of Cooper pairing by the local
Zeeman field. The order parameter also exhibits two maxima near the edge of
the Fermi cloud, which is an unique feature in the 1D case \cite{liu07}.
Remarkably, the phase of the order parameter changes linearly (the property
of the FF type of order parameter) inside the barrier and remains constant
outside as shown Fig.~\ref{f1}(c). The constant values of the phases are
different on the left and right sides of the barrier and can be any value,
demonstrating a JJ with tunable phase. In a 2D system separated by a
magnetic barrier chain, such phase junction still exists.

\begin{figure}[tbp]
\includegraphics[width=3.4in]{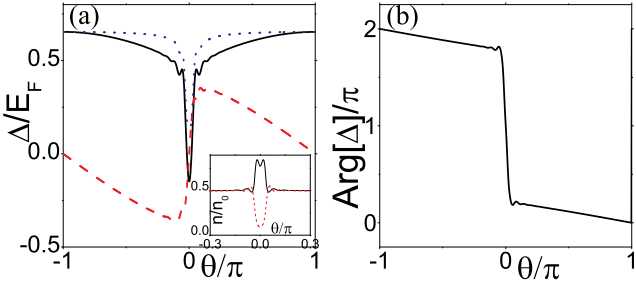} 
\caption{(Color online) Phase tunable JJ in a ring. (a) The real part (black
solid), imaginary part (red dash), and absolute value of the order parameter
(blue dotted). Inset: Density profiles for $|\uparrow >$ state (black solid)
and $|\downarrow >$ state (red dash). (b) The phase of the order parameter.
In numerical calculation, we take the radius $R=1$, $\hbar =1$. The
non-interacting Fermi momentum is defined as $k_{F}=\protect\pi n_{0}/2$ and
the Fermi energy $E_{F}=\hbar ^{2}k_{F}^{2}/2m$ with $n_{0}=N/2\protect\pi R$
being the average density. The parameters are $\protect\lambda %
=2.3E_{F}/k_{F}$, $\protect\gamma =2.4$, $\Omega _{R}=0.8E_{F}$, $\protect%
\delta =0.25E_{F}$, $u_{mag}=1.5E_{F}$, $a_{mag}=0.1 $, $\protect\mu %
=-0.41E_{F}$ and $N=60$.}
\label{f3}
\end{figure}

Such tunable phase across the magnetic barrier also exists inside a ring
trap. However, the phase outside the barrier is not constant anymore due to
the periodic confinement. To clearly distinguish the phase change in the
barrier from the bulk region, we consider a rectangular-shaped barrier
located at $\theta =0$, described by $V_{Z}(\theta )=-u_{mag}[\Theta (\theta
+a_{mag})-\Theta (\theta -a_{mag})${]} with the step function $\Theta (x)$.
We have studied realistic Gaussian potentials and the results are
qualitatively the same with a small quantitative deviation from the
rectangular one. In Fig.~\ref{f3}, we present our self-consistent BdG
results. Similar as the harmonic trap case, in Fig.~\ref{f3}(a), there is a
local imbalanced region inside the barrier (the inset figure) where both
real and imaginary parts of the order parameter are nonzero and have
different structures. More interestingly, as shown in Fig.~\ref{f3}(b), the
phase of the order parameter outside the barrier changes almost linearly,
instead of constant, due to the periodic boundary condition.

\begin{figure}[tbp]
\includegraphics[width=0.48\textwidth]{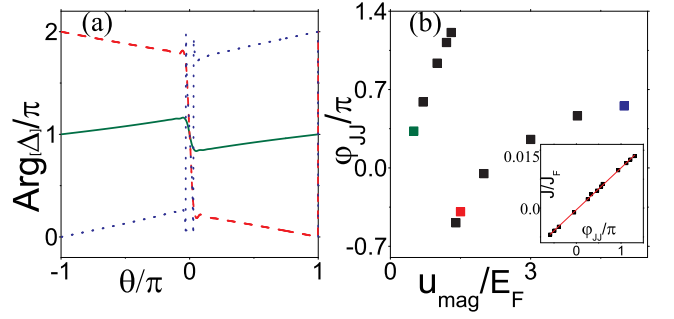} 
\caption{(Color online) (a) The phase structure of the order parameter in
the real space with different barrier strengths. (b). Tunable phase of the
JJ vs. barrier strength. The three colored squares correspond to the three
cases in (a) respectively. Inset: Spontaneous mass current as the function
of the phase of the JJ. Mass current in unit $J_{F}=n_{0}v_{F}$ where $v_{F}$
is the Fermi velocity $v_{F}=\hbar k_{F}/m$.}
\label{f4}
\end{figure}

The phase difference across the JJ can be continuously tuned. In Fig.~\ref%
{f4}(a), we present three different phase structures for three different
barrier strengths $u_{mag}$. To quantitatively characterize the phase
change, we introduce two types of phase differences: $\varphi _{dif}=\phi
_{L}-\phi _{R}$ between two edges of the barrier and the phase change $%
\varphi _{JJ}$ outside the barrier. To avoid the complication from the
periodicity of the phase angle, we treat the order parameter phase in its
principal value of $[0,2\pi )$. With increasing barrier strength, $\varphi
_{dif}$ increases from less than $\pi $ (green solid) to larger than $\pi $
but smaller than $2\pi $ (red dashed), and finally to even larger than $2\pi
$ (blue dotted). Obviously, for small barrier strengths, $\varphi
_{JJ}=\varphi _{dif}>0$ (green solid); for moderate ones, $\varphi
_{JJ}=\varphi _{dif}-2\pi <0$ (red dashed); for strong ones, $\varphi
_{JJ}=\varphi _{dif}-2\pi >0$ (blue dotted). In Fig.~\ref{f4}(b), we plot $%
\varphi _{JJ}$ as a function of the magnetic barrier strength. $\varphi _{JJ}
$ has a discontinuous point near $\varphi _{JJ}=1.2\pi $. In fact, when $%
\varphi _{dif}$ is around $\pi $, there are two steady states corresponding
to positive and negative $\varphi _{JJ}$, respectively. When $\varphi
_{dif}>1.2\pi $, the state with negative $\varphi _{JJ}$ has lower energy
and becomes the ground state. We note that the phase difference of the JJ
can also be tuned by changing other parameters such as the barrier width and
the atom-atom interaction strength.

\emph{Spontaneous mass current:} The linear phase gradient outside the
barrier of the JJ induces a spontaneous mass current in the ring trap, which
is defined as
\begin{equation}
J(\theta )=\frac{\hbar }{mR}\underset{\sigma =\uparrow ,\downarrow }{\sum }%
\text{Re}\langle \psi _{\sigma }^{\dagger }(\theta )(-i\partial _{\theta
}+L_{\sigma })\psi _{\sigma }(\theta )\rangle .  \label{eq:current}
\end{equation}%
Here $L_{\uparrow }=-L_{\downarrow }=L$ and the second term originates from
the SOAM coupling. The mass current is zero for the linear 1D system in the
harmonic trap, where there is no phase gradient outside the barrier. The
inset of Fig.~\ref{f4}(b) shows that the ground state of JJ with finite $%
\varphi _{JJ}$ exhibits a finite mass current, which is linearly
proportional to $\varphi _{JJ}$. Such magnetic barrier induced mass current
demonstrates the magnetoelectric effect for cold-atoms, which may have
important applications in atomtronics.

\begin{figure}[tbp]
\includegraphics[width=3.4in]{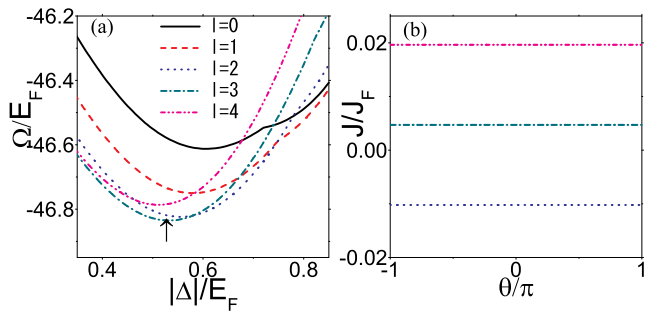} 
\caption{(Color online) Ring structure with uniform two-photon detuning. (a)
The thermodynamic potential as a function of $|\Delta |$, with different
angular momentum of the phase. The arrow shows the ground state. The
parameters are $u_{mag}=0$, $\protect\mu =-0.435E_{F}$, the other parameters
are the same as Fig.~\protect\ref{f3}. (b) Spontaneous mass current for
different $l$.}
\label{f2}
\end{figure}

The tunable phase across the barrier and the spontaneous mass current along
the ring could be understood from the FF order parameter of the superfluid
in the presence of SOAM coupling and in-plane Zeeman field. Here we
illustrate this mechanism by considering a uniform in-plane Zeeman field
along the whole ring (i.e., $V_{Z}(\theta )=\delta $). We find the ground
state of the system possesses finite angular momenta Cooper pairs $\Delta
(\theta )=\Delta _{0}\exp (il\theta )$ with $\Delta _{0}$ being constant and
$l$ being an integer due to the periodic boundary condition. To obtain the
order parameter numerically, we start from random initial order parameters
and then self-consistently solve the BdG equation in the real space until it
converges. We find that each converged final state always corresponds to a
state with certain $l$, suggesting that these states are steady states. To
see this more clearly, we choose $\Delta (\theta )=\Delta _{0}\exp (il\theta
)$ and compute the thermodynamic potential for each $l$ with fixed chemical
potential as a function of $|\Delta (\theta )|$. In the mean-field theory,
the thermodynamic potential $\Omega $ is defined as $\Omega =\langle
H\rangle -\int Rd\theta \,|\Delta (\theta )|^{2}/g_{1D}$, which can be
expanded in the angular momentum space (similar to the momentum space in a
traditional homogeneous 1D system). In Fig.~\ref{f2}(a), we plot the
thermodynamic potential with respect to $|\Delta (\theta )|$ for different $l
$, showing that there always exists a local minimum of the thermodynamic
potential for each $l$. We find that the converged state obtained by the
self-consistent calculation in the real space exactly corresponds to the
local minimum for each $l$. The ground state is the one with the lowest
thermodynamic potential and $l=3$ for the particular parameters shown in
Fig.~\ref{f2}(a). With the local magnetic barrier, the FF phase $\exp
(il\theta )$ changes across the barrier, leading to the tunable phase
junction.

For the FF states in an infinite 1D homogeneous system, the mass current $%
J\propto \partial \Omega (Q)/\partial Q|_{Q=Q_{0}}$ with $\Omega (Q)$ being
the thermodynamic potential for $\Delta (x)=\Delta _{0}e^{iQx}$ and $Q_{0}$
being the center-of-mass momenta of Cooper pairs of the ground states. This
current equals to zero because $\partial \Omega /\partial Q|_{Q=Q_{0}}=0$ is
satisfied for the ground states~\cite{ff64,qu13}. However, in the
ring-shaped system, instead of taking continuous values, $l=QR$ can only be
discrete due to the periodic boundary condition (note that the superposition
of different $l$ states is not energetically preferred), therefore $\partial
\Omega (Q)/\partial Q|_{Q=l/R}$ can be nonzero for a finite ring-shaped
system (i.e., $1/R\neq 0$), leading to finite mass current for the ground
states. The direction of the current is also dictated by the sign of $%
\partial \Omega (Q)/\partial Q|_{Q=l/R}$. In Fig.~\ref{f2}(b), we plot the
mass currents for three steady states corresponding to different values of $%
l $. They are all nonzero. The current of the ground state is smaller than
other states with different $l$ because of smaller $\partial \Omega
(Q)/\partial Q|_{Q=l/R}$.

\emph{Experimental realization and observation}: In experiments, we consider
$^{40}$K atoms and utilize LG laser beams to generate a ring trap as well as
the SOAM coupling between two hyperfine states \cite{Hu15,pu15,sun15,qu15}.
The magnetic barrier can be generated by a tightly focused laser beam \cite%
{bloch11}. When the wavelength of the focused laser lies between D1 and D2
transition lines, atoms at different hyperfine states experience different
potentials, leading to spin-dependent potential. To measure the current, one
can sample one slice of the Fermi ring and measure its momentum distribution
\cite{Zwierlein14}. In this slice, the momentum difference of the atom cloud
between the tangential direction of the ring and the opposite direction
determines the local current. In addition, one can consider the Doppler
induced interference of the phonon modes, which has been utilized to measure
the current in toroidal BECs \cite{Stringari15}. Finally, because the
mechanism for generating FF order parameters in the magnetic barrier are the
same for 1D, 2D and 3D \cite{zheng13,hu13}, the proposed phase tunable JJ
should also apply to a 2D spin-orbit coupled Fermi gas with a magnetic
barrier line or 3D with a magnetic barrier plane. In a ring trap, this means
the radial confinement need not be very strong, corresponding to a 3D
toroidal trap.

\emph{Summary}: In summary, we propose that the ground state of a Josephson
junction with arbitrary and tunable phase can be realized in spin-momentum
coupled Fermi superfluids in a harmonic trap or SOAM coupled Fermi
superfluids in a toroidal-shaped trap. When a different phase from the
ground state value is applied externally, it is known that a finite
Josephson current is generated. We find that a spontaneous mass current
exists in a finite ring-shaped system due to the periodic boundary
condition, demonstrating the magnetoelectric effects in cold atoms. The
experimental realization of such tunable phase JJ may open novel
possibilities for many applications in atomtronics, such as superfluid phase
battery and rectifiers, phase-based quantum bits, and the observation of
topological superfluids and the associated Majorana fermions.

\emph{Acknowledgements: }We thank C. Wu for helpful discussion. This work is
supported by ARO (W911NF-12-1-0334), and NSF (PHY-1505496).

\end{document}